# Sungrazer Comet C/2012 S1 (ISON)

## Curve of light, nucleus size, rotation and peculiar structures in the coma and tail

*T. Scarmato[1]*

[1]Toni Scarmato's Observatory, via G. Garibaldi 46, 89817 San Costantino di Briatico, Calabria, Italy

Last revision 2014 May 13

## Abstract

We present our results for comet (ISON). **C/2012 S1 (ISON),** a sungrazing comet discovered on the 21 September, 2012 by Vitali Nevski (Виталий Невский, Vitebsk, Belarus) and Artyom Novichonok (Артём Новичонок, Kondopoga, Russia), using the 0.4-meter (16 in) reflector of the International Scientific Optical Network (ISON) near Kislovodsk, Russia and the automated asteroids-discovery program CoLiTec. Comet C/2012 S1 had has the closest approach to the Sun on 28 November, 2013 at a distance of 0.012 AU. The solar radius measure 695,500 km so ISON passed around $1.1 \times 10^6$ km above the Sun's surface. The orbital parameters based on a very long time's period, give an eccentricity $e > 1$ so the orbit is hyperbolic. This is in agreement with a dynamically new comet coming from the Oort cloud. Other important closest approaches, were from Mars on 1 October, 2013 at about 0.0724 AU and from Earth on 26 December, 2013 at about 0.429 AU.

Our results were obtained with amateur telescopes observing from the ground, before and after the comet had passed conjunction with the Sun in August 2013. We measured a mean radius of 830 +/- 245 m, and we have identified a persistent structure in the direction of the sun with a preliminary rotation period of approximately 3 days. In the tail are visible peculiar structures linked to the production of dust by the nucleus. Our data Af(rho) (A'Hearn et al. 1984), a physical parameter that indicates the amount of dust produced, is approximately 500+/-50 cm in the quite phase which means a production $Q_{dust}=500$ kg/sec equal to $43.2 \times 10^6$ kg per day. A considerable amount for a small comet. We have also detected two major outbursts in January, 2013 and November, 2013, and minor outbursts throughout the observation period.

## 1) INTRODUCTION

Most of our observations were made with amateurs instruments between September 2012 and November 2013. We added also some observations made with a 2 meter telescope of Las Cumbres Organization. In the table below are the observers involved in this work.

| Name | Location | Telescope | CCD | Filter |
|---|---|---|---|---|
| Toni Scarmato | Toni Scarmato's Observatory, San Costantino di Briatico, Calabria, Italy | 25 cm Newton | Atik 16 IC mono | Rc (Schuler) |
| Efrain Morales | Observatory, Aguadilla, Puerto Rico | 30 cm S.C. | SBIG ST-402me | LIGHT |
| Nick Howes | Las Cumbres – LCOGT | LCOGT-FTN 200 cm | FI CCD486 BI DB | R (Bessel) |
| Antonino Brosio | AOB Observatory, Rosarno, Calabria, Italy | 25 cm Newton | Atik 314 L mono | Rc (Schuler) |
| Bruce Gary | Hereford Arizona Observatory | 35 cm S.C. | SBIG ST-10 | Cb (LIGHT) |
| Dennis Whitmer | Private Observatory | 35 cm S.C. | SBIG ST-1301 | Rc (Bessel) |
| Gustavo Muler | Canariae (Lanzarote) | 30 cm S.C. | SBIG ST-8XME | LIGHT |
| J.A. Jimenez | Observatorio Astronómico de Yunquera | 25 cm S.C. | SBIG ST-7 | CLEAR |

For details regarding some observer's CCDs and the bandpass of the filters used, go to the Appendix 1.

All the images were calibrated with dark, flat and bias if needed. The steps to obtain our results are based on the two main procedures; measuring the magnitude and Af(rho) in R band, and elaborating the images with mathematical algorithms to measure the size of the nucleus, detect variations of position of structures in the coma and to analyze the peculiar structures in the tail. When the images were taken with other filters or without a filter, we used the equation of conversion of the magnitude in the R band equivalent. Our main goal, other than the construction of the Curve of Light before the perihelion that permit analysis of the photometric behavior of the comet linked to its activity, is to demonstrate that also with amateurs instruments it is possible to measure the size of a comet using professional astronomers procedure up to now applied only to Hubble Space Telescope images.

That procedure was applied to images at a lower resolution with respect Hubble images, and the results are in agreement with professional results. We don't want to obtain very precise measurement of the size of the nucleus; in fact we have computed an error of about 20% of the measure, but our goal is to obtain a measurement that is at least of the same order as professionals measurements. The results obtained are very good if we consider that, for example, we obtained better measurements of the radius of ISON in agreement to that of professional astronomers with a **25 cm Newton Telescope Reflector** and a CCD with a resolution of 1.27 arcsec/pixel at about 2.9 A.U.. This means that the spatial resolution of our original fits images is about 2700 km at that distance. The bigger nucleus of a comet observed just now was Hale-Bopp nucleus esteemed about 40 km in diameter. So we can say that our work was really a challenge if we consider that the first estimations of the nucleus of ISON, (Central Bureau Electronic Telegrams, 3496,1 (2013). Edited by Green, D.W.E.), with a preliminary analysis by professional astronomers with images of ISON taken on 2013 April 10 by Hubble Space Telescope, using a coma-nucleus separation technique **(cf. Lamy et al. 2009, A. Ap. 508, 1045),** suggests a nuclear radius less than 2 km. **(C/2012 S1 (ISON) Jian, Y. Li et al. 2013).** Our technique is based on the same procedure implemented with other powerful algorithms.

In the next paragraphs, we describe the steps and algorithms used in our work, and, in particular, we will describe the procedure to separate the nucleus from the coma and to measure the contribution of the brightness of the nucleus with respect the brightness of the coma and how to compute the residuals in ADU **(Analog/Digital Unit)** of the central brighter pixel that permit us to compute the apparent magnitude and so the absolute magnitude of the nucleus and, from this, the radius in meters.

## 2) METHODOLOGY

### 2.1   Photometry of aperture

Our methodology is based on the comparison of the flow of the object under investigation with the flows of stars that are in the same field. For comets, images should be made preferably with the photometric R Johnson-Cousin filter, the band in which the dust in the coma of the comet reflects the light coming from the Sun. The series should be as long as possible, at least two hours along with exposure times to have the S/N highest possible but taking into account that the comet has moved in the individual exposure. If the telescope is on an equatorial mount, it must be well aligned with the pole and problems bending and tracing must to be minimized. We stress, however, that the images with objects can be moved and measured as the total flow is always captured by the pixels of the CCD. It's important not to include windows opening so as to contain the flow, but not too large to incorporate any background stars near the source.

The images will then be calibrated with the masterdark, masterflat and if necessary with the masterbias. If it is possible, we must maintain a stable temperature of the CCD by at least in +/- 0.5 °C. If we are unsure of what is most important to do a series of darks during the observation in order to mediate all the images and obtain a master dark refers to all possible temperatures occurred throughout the observing session. Depending on the brightness of the comet, the 'quantum efficiency of the CCD and taking into account that the R photometric filters reduces the total quantity of photons arriving in the CCD compared with images unfiltered, unless you have a formula to get the appropriate time optimum exposure, is bypassing the problem of doing tests with increasing exposure times from at least 60 seconds. When the counting of ADU (**Analog/Digital Unit**) is equal to 70% of the level of saturation of the CCD, then you can proceed to capture the full range of images. To have a better S/N (Signal/Noise), it must be taken into account that the motion of the comet does not exceed 90% of the level of saturation of the CCD camera. Once the images are made to their development, first of all calibrating them then with a program that can measure the ADU of individual image automatically, you can get the count for each image. In our case, was used IRIS. The program, once having measured the individual sources in the single image, provided an output file with dates in JD – ADU and the values of the objects for each image recorded in chronological order. (**see Table 1**)

Tab. 1 – sample records of counts with the fluxes measured in ADU; the first column shows the Julian Data of the observation for each image; in subsequent columns the counts of the object that you want the photometry and stars reference are shown.

| JD | FLUXCOMET | REF1 | REF2 |
|---|---|---|---|
| 2454718.3129500 | 21304 | 510508 | 44780 |
| 2454718.3136991 | 22548 | 521433 | 45027 |
| 2454718.3144502 | 22145 | 518231 | 39568 |
| 2454718.3152014 | 23422 | 515335 | 42985 |
| 2454718.3159549 | 22926 | 509435 | 42423 |
| 2454718.3167083 | 22180 | 512611 | 37962 |

Since the program used can do photometry automatically on only 5 stars at a time, if you need more than 5 photometric objects, you will need more than one file with the data of individual objects. To put them together, you can use Excel and then create a single file txt type to be able to develop a program of analysis of the curves of light. In our case we used GNUPlot and Excel. It then proceeds with the comparison between the flow of the object under investigation and the flow of the other stars measured in the field. The formula

$$Flux = \frac{AduStar}{\sum Adu(ref1, ref2, ref3, ....)} \quad (Eq.\ 1)$$

permits to obtaining the values that are fitted by a linear function that determines the best fit of points of the curve of light. The sum in the denominator to obtain a theoretical star whose flow is equal to the sum of the flows of individual stars. In this way, you can build a chart that, if precision is desired, highlights the brightness variations that appear as positive or negative peaks in the curve of light.

Table 2

| Star/Catalog | mag. - USNO B1 | mag. - TYCHO 2 | mag. - SDSS |
|---|---|---|---|
| T 2574:136:1 (Ref1) | R1=8.95 – R2=8.89 | Bt=10.946 – Vt=10,629 | R=9.340 |
| T 2574:578:1 (Ref2) | R1=9.77 – R2=9.69 | Bt=12.147 – Vt=10,629 | R=10.119 |
| Example of photometric parameters of the stars of reference derived from catalogues online using Aladin tool, after having calibrated the image and aligned on the stars. | | | |

Using formulas, **magstar = - 2.5 * LOG ((1 + (aduref2 + aduref3) / aduref1)) + magref1**, where magref1 R is the magnitude of the star brighter used as a reference, obtained from USNOB1 catalogue, (see Table 2), we derive the magnitude of the star of reference. Through the formula **magcomet = - 2.5 * LOG (aducomet/(aduref1 + aduref2 + aduref3)) + magstar** we have the magnitude of the comet. To compute the error, we used the standard deviation at 1 sigma of precision.

## 2.2 Nucleus size estimation

The study of the size of cometary nuclei and of their physical properties, is important to understand the formation and evolution of the Solar System. Up to this time, the determination of the nucleus size of a comet, was made only with Hubble telescope images. The camera used with HST has a resolution of 0.04 arcsec/pixels. This means that a comet observed at 4 A.U. (1 A.U. is 149.577.000 km) permit to reach a spatial resolution of about 120 km. Because the cometary nuclei are smaller than spatial resolution, one needs to use the algorithms to separate the contribution of the nucleus from the coma, to the total brightness coma + nucleus. **(Lamy et al. 2009)**.

We have developed one method to measure the nucleus of a comet also with amateur images at lower resolution (see Table 1). The goal is to measure the ADU for the nucleus and using the stars in the FOV of the images and the R magnitude to compute the radius of the comet. With the formulae in **Lamy et al. 2009**, we are able to measure the nucleus size of a comet to within about 20%. Using the following model of the luminosity

$$L(\rho) = [k_n \delta(\rho) + coma] \otimes PSF \quad (Eq.\ 2)$$

with a model of the comet

$$Model = [nucleus + coma] \otimes PSF \quad \textit{(Eq. 3)}$$

where PSF denotes the Point Spread Function of the telescope and $\otimes$ represents the Convolution operator. To resolve the nucleus, in the sense that it is possible to measure the spatial size, we can use the following equation;

$$nucleus = k_n\ \delta(\rho) \quad (Eq.\ 4)$$

where δ is the Dirac's delta function, ρ is the radial distance from the center, and $k_n$ is a scaling factor. The apparent magnitude m in R band of a comet can to be linked to its physical properties by the equation

$$p\Phi(\alpha)R_n^2 = 2.238 \cdot 10^{22} r_{sc}^2 \Delta^2 10^{0.4(m_s - H)} \quad (Eq.\ 5)$$

where **p** is the geometric albedo, $\Phi(\alpha)$ the phase function, $R_n$ is the radius of the comet, $r_{sc}$ the Sun-Comet distance, Δ the Earth-Comet distance, $m_s$ the absolute Sun's magnitude in band R=-27.09 and H the absolute nucleus magnitude in R band. With simple transformations, we can obtain the radius of the comet in meters;

$$R_n = \frac{1.496 \, 10^{11}}{\sqrt{p}} 10^{0.2(m_s - H)} \quad \text{(Eq. 6)}$$

where

$$H = m - 5\log(r_{sc})\Delta - \alpha\beta \quad \text{(Eq. 7)}$$

$$\alpha\beta = -2.5\log[\Phi(\alpha)] \quad \text{(Eq. 8)}$$

where $\alpha$ is the phase angle, m is the apparent magnitude of the nucleus measured from the observation in R band and β the phase coefficient. We assumed $r_{sc}$=1 A.U., β=0.04 mag/deg and p=0.04 as standard values.

## 3) PROCESSING IMAGES

### 3.1    1/rho model and background subtraction

After having calibrated the original image with dark, flat and bias, we applied an algorithm based on Bonev, T. & Jockers, K. "Spatial distribution of the dust color in comet C/LINEAR (2000 WM1)" (Proceedings of Asteroids, Comets, Meteors - ACM 2002. International Conference, 29 July - 2 August 2002, Berlin, Germany. Ed. Barbara Warmbein. ESA SP-500. Noordwijk, Netherlands: ESA Publications Division, ISBN 92-9092-810-7, 2002, p. 587 – 591). This algorithm  extracts the pixels value of the coma, subtracts the background value and multiplies for the cometcenter distance to create a new image with the computed values.  After this, we used a crop of 40x40 pixels image centered on nucleus position, to apply the below procedures (Bicubic Interpolation, Convolution and PSF).  With this method, we obtained results in agreement with professionals astronomers, using Hubble Space Telescope images taken on April and May 2013, published in

(Comet C/2012 S1 (Ison)

Jian,Y.Li; Weaver, H. A.; Kelley, M. S.; Farnham, T. L.; A'Hearn, M. F.; Knight, M. M.;

Mutchler, M. J.; Lamy, P.; Toth, I.; Yoshimoto, K.; Gonzalez, J. J.; Shurpakov, S.; Pilz, U.; Scarmato, T.

Central Bureau Electronic Telegrams, 3496, 1 (2013). Edited by Green, D. W. E. )

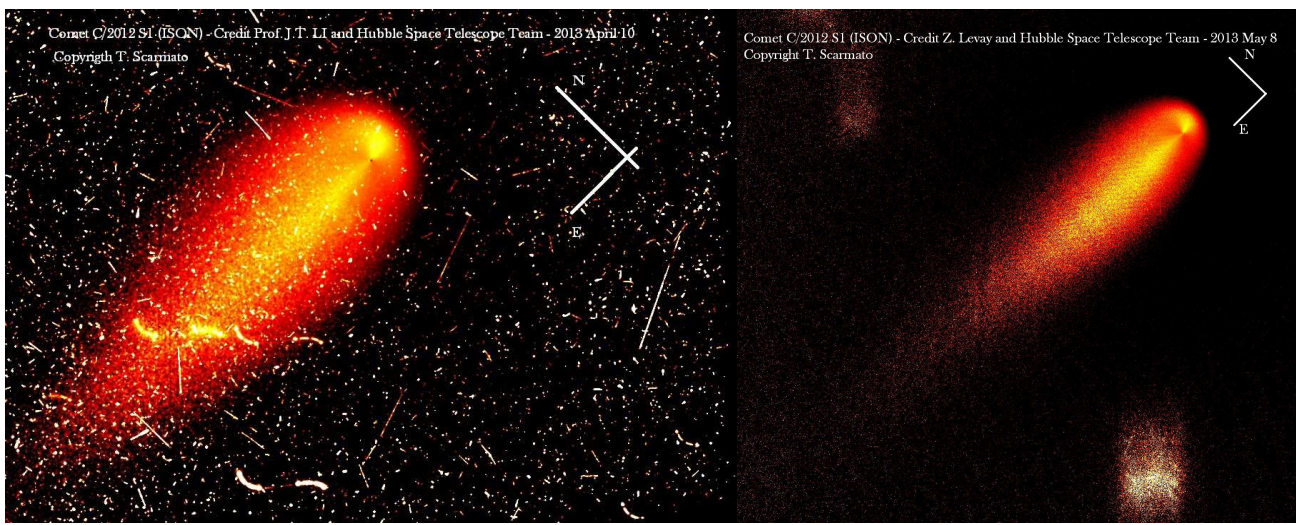

Comet C/2012 S1 (ISON)

**Left:** Hubble Space Telescope image Credit J.Y. Li et al.  **Right:** Hubble Space Telescope image Credit Z. Levay et al.

**Elaboration T. Scarmato** using a 1/rho Model applied based on procedure in:

Bonev, T. & Jockers, K. "Spatial distribution of the dust color in comet C/LINEAR (2000 WM1)

## 3.2 Bicubic Interpolation

In mathematics, **Bicubic Interpolation** is an implementation of cubic interpolation for interpolating data points on a two directions x and y of a regular image pixles. It preserves fine details better than the common bilinear algorithm.
http://en.wikipedia.org/wiki/Bicubic_interpolation

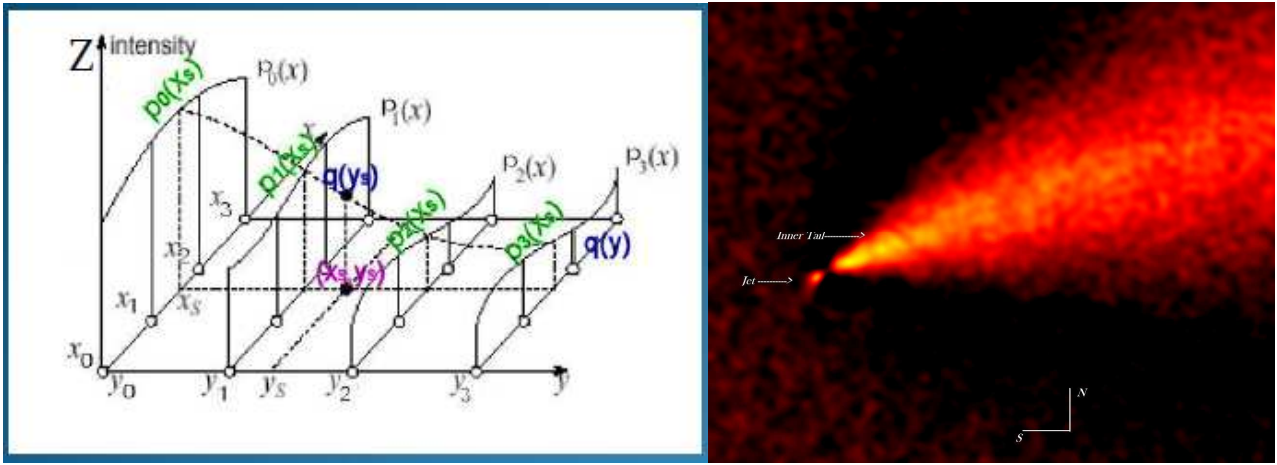

**Fig.1-Left**: Bicubic Interpolation along X and Y axis 3x3 resampling **Right:** 4x4 Bicubic Interpolation+ Convolution+ PSF. Bicubic Interpolation works on the 16 pixels around the position for which we want the value and permit to obtain the finer structure of the image. On our image we used 4x4 resampling to divide one pixel in 16 sub-pixels. The Convolution+ PSF correct the problems due to focus, aberration bad tracking etc.. Finally the LS and $1/\mathrm{rho}^n$ model filters work on the sub-pixels and are able to identify the finer **GRADIENT OF BRIGHTNESS** in the coma and tail.

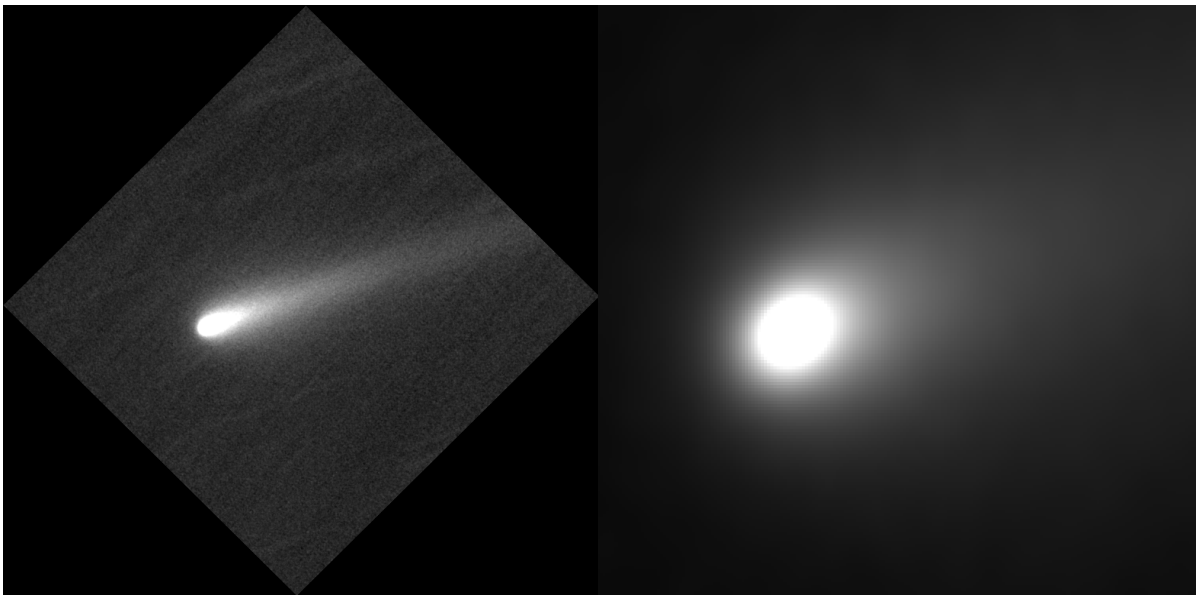

**Fig. 2 - Left**: Original image oriented taken on 2013 October 15 - **Right**: Pixellization, resampled 4x4 Bicubic Interpolation + Convolution + PSF

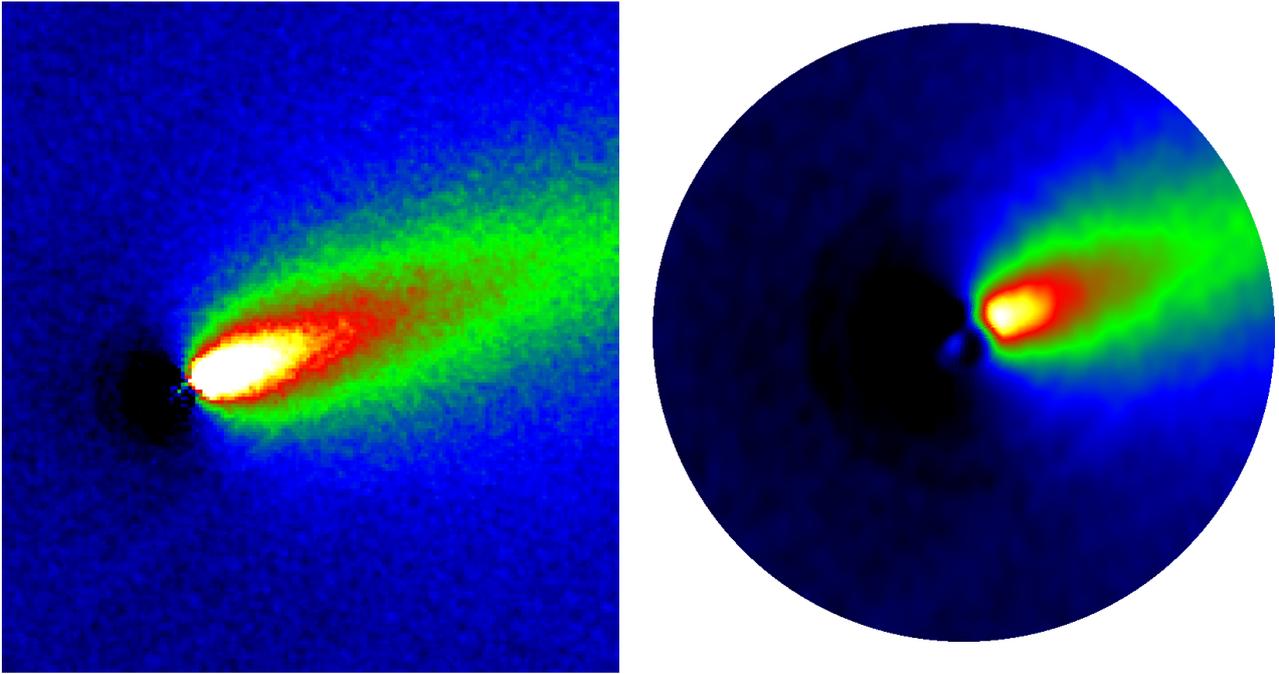

**Fig. 3** – **Left**: $1/\text{rho}^n$ model Filter on image unresampled – **Right**: $1/\text{rho}^n$ model Filter on image resampled

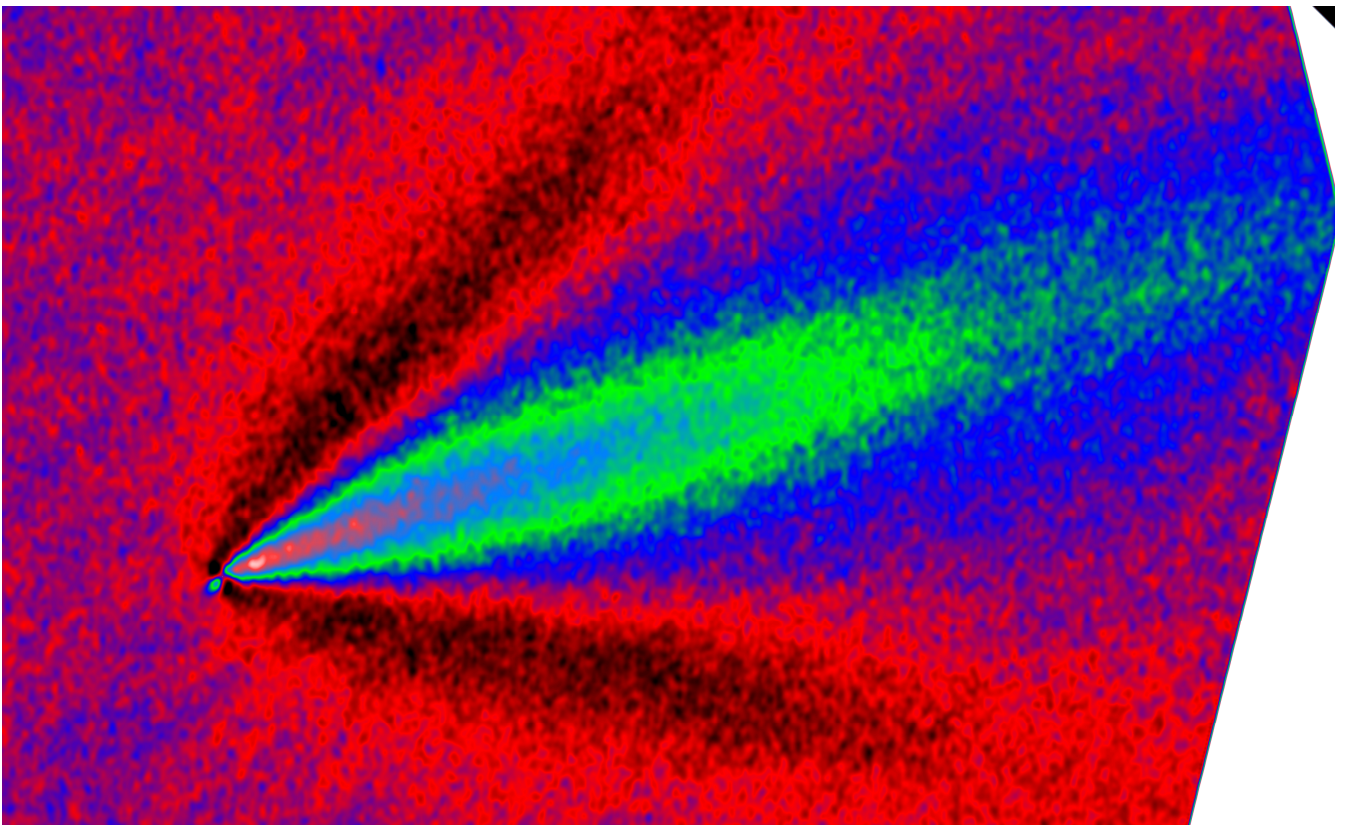

**Fig. 4** – Larson-Sekanina Filter applied to the Resampled 4x4 Bicubic Interpolation + Convolution + PSF

Algorithm of the BICUBIC INTERPOLATION with MATLAB on 40x40 pixels image

```
>> src= imread('soccer_src.fits');
%Construction of two matrixs x,y from the vector [1:40] abscissa (x) and ordinate (y) of the point P(x,y) in the grid.
>> [x,y] = meshgrid(1:40);
>> [xi,yi] = meshgrid(1:0.50:40);
>> fork = 1:3
soccer_cub(:,:,k) = interp2(x,y,double(src(:,:,k)),xi,yi,'bicubic');
>> end
>> subplot(1,2,1);
>> image(src);
% the two axis have the same lenght
>> axissquare;
>> subplot(1,2,2);
>> image(uint8(soccer_cub));
>> axissquare;
```

This procedure permits the construction of a finer structure of the coma and to define the values of the sub-pixels around the cometcenter pixel that contain the nucleus. In this manner, we can obtain details of the coma structure and the residuals between the brighter pixel and the other around with the Lamy et al. separation procedure describe above. The Dirac functions permits the interpolation of the value of the sub-pixels computed with the bicubic interpolation. **(R. Keys, 1981).**

### 3.4 Convolution

The convolution matrix in computer graphics is used to apply filters to images. Through the application of the convolution of two two-dimensional arrays of which the first is the original image and the second, called the kernel, is the filter to apply. Consider the matrix A that represents the matrix containing the gray values of all the pixels of the original image and the matrix B that represents the kernel matrix. Superimpose the array B to array A so that the center of the matrix B is in correspondence with the pixels of the matrix A to be processed.

The value of each pixel of the matrix A processing object is recalculated as the sum of the products of each element of the matrix kernel with the corresponding pixels of the matrix A below. So we can to define the convolution of *f* and *g*, written $f \otimes g$, as the integral of the product of the two functions after one is reversed and shifted. Mathematically we have:

$$(f * g)(x) = \int_{-\infty}^{+\infty} f(y)g(x-y)dy = \int_{-\infty}^{+\infty} f(x-y)g(y)dy \quad \text{(Eq. 9)}$$

## 3.5 PSF (Point Spread Function)

The **point spread function** (**PSF**) is strictly linked with the response of a telescope to the light of a **point source** or point object, so the PSF is the response of a telescope in focus when observe the light of an object affected by some variable. In our context, when we observe astronomical object from the ground, the photons that impact with CCD don't produce a point of light but an extended blob that represents an unresolved object. So the image of an astronomical object can then be represented as a **convolution+PSF** of the real object in the sky. Considering the linearity property of a telescope,

$$Pixel(Obj_1 + Obj_2) = \text{Pixel}(Obj_1) + \text{Pixel}(Obj_2) \quad \text{(Eq. 10)}$$

the image of an object is linked with the *superposition principle* for linear systems. The images obtained are called **point spread functions**, that represent the fact that a photon that impact with a point of the area of the CCD produce the *spread* out of the light to form a finite area in the image and not a single point. Because an astronomical object that emit light can to represented as discrete point objects with different intensity, the image of the object in the telescope is computed as a sum of the PSF of all the single points, so is need to know the optical properties of the telescope. To do this we used a **convolution equation**. In astronomy, knowing the PSF of the CCD+Telescope+Filters system is very important for reconstruct the real image. For ground based optical telescopes, the seeing is dominating the PSF. Relevant is the fact that in high-resolution ground-based telescope, the PSF is subject to the effect called anisoplanatism.

## 3.6 PSNR AND MSE

**Peak signal-to-noise ratio**, often abbreviated **PSNR**, is the computation of the ration between the maximum possible intensity of a pixel/value the intensity of the noise/value in the image. Because many images have a very wide dynamic range as the astronomical CCD images, to represent the PSNR is using the logarithmic.

PSNR can to be defined using the mean squared error (*MSE*). So, given a noise-free two dimensional axb monochrome image A with its noise N, we have the following equation:

$$MSE = \frac{1}{ab} \sum_{i=0}^{a-1} \sum_{j=0}^{b-1} [A(i,j) - N(i,j)]^2 \quad \text{(Eq. 11)}$$

# 4) RESULTS

## 4.1 Curve of light in R band

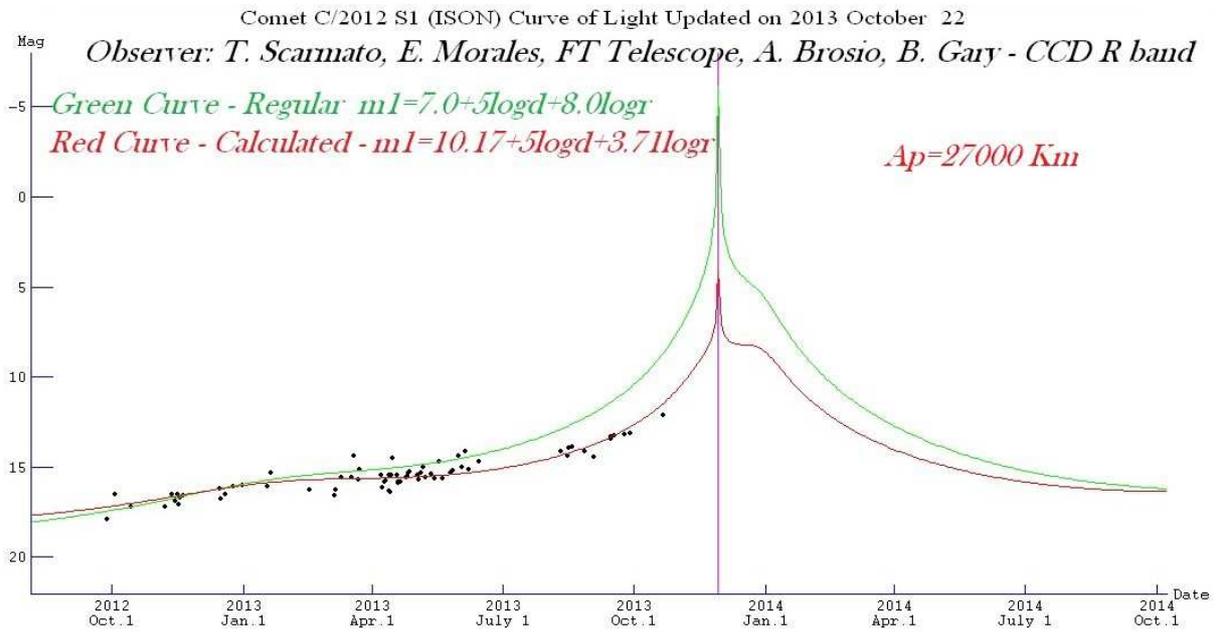

## 4.2 Phase's effects correction

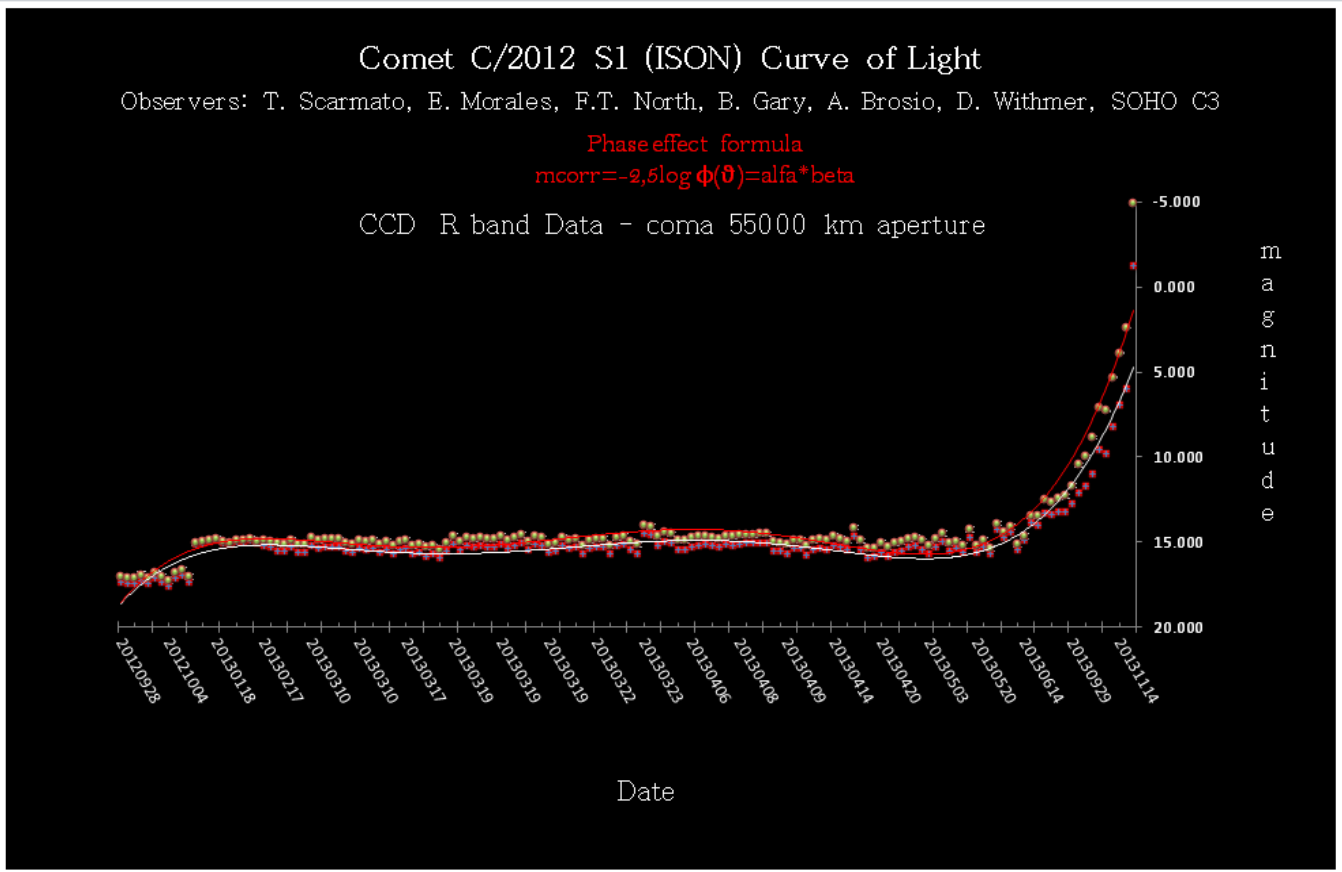

## 4.3 Af(rho) and outbursts

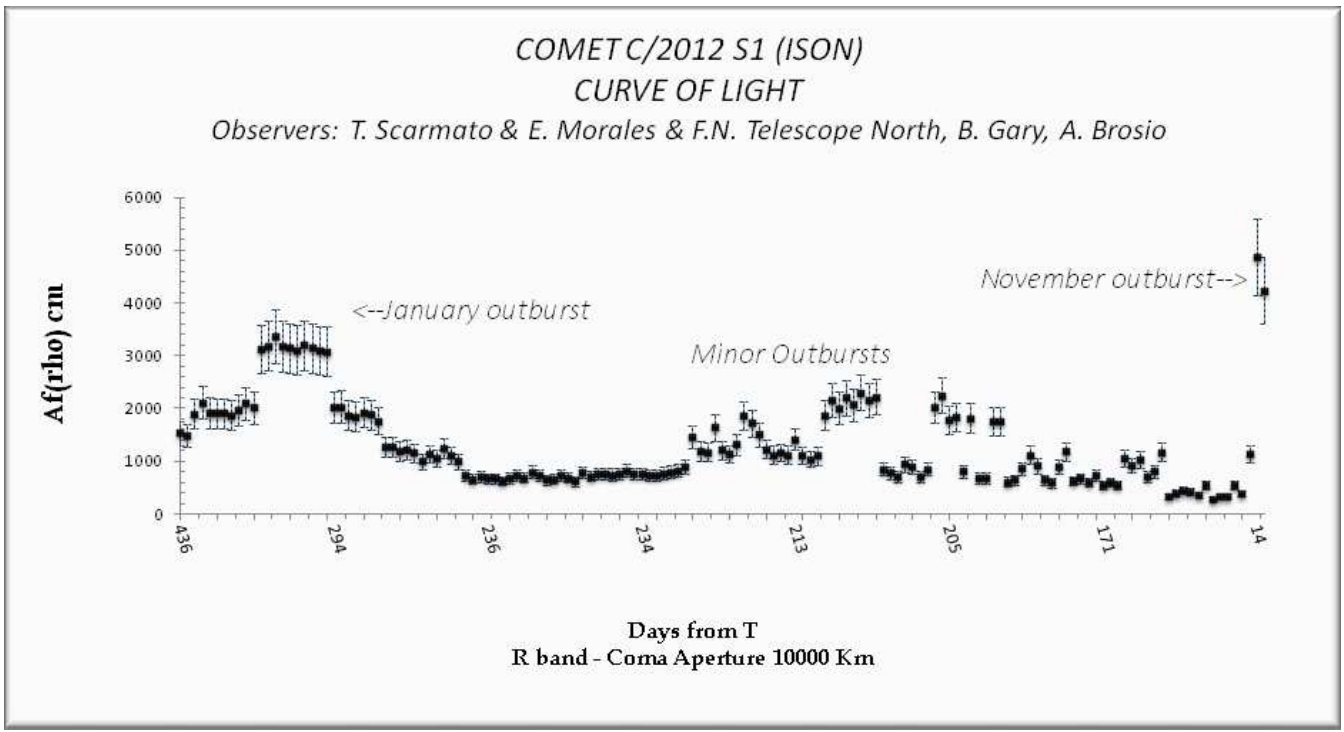

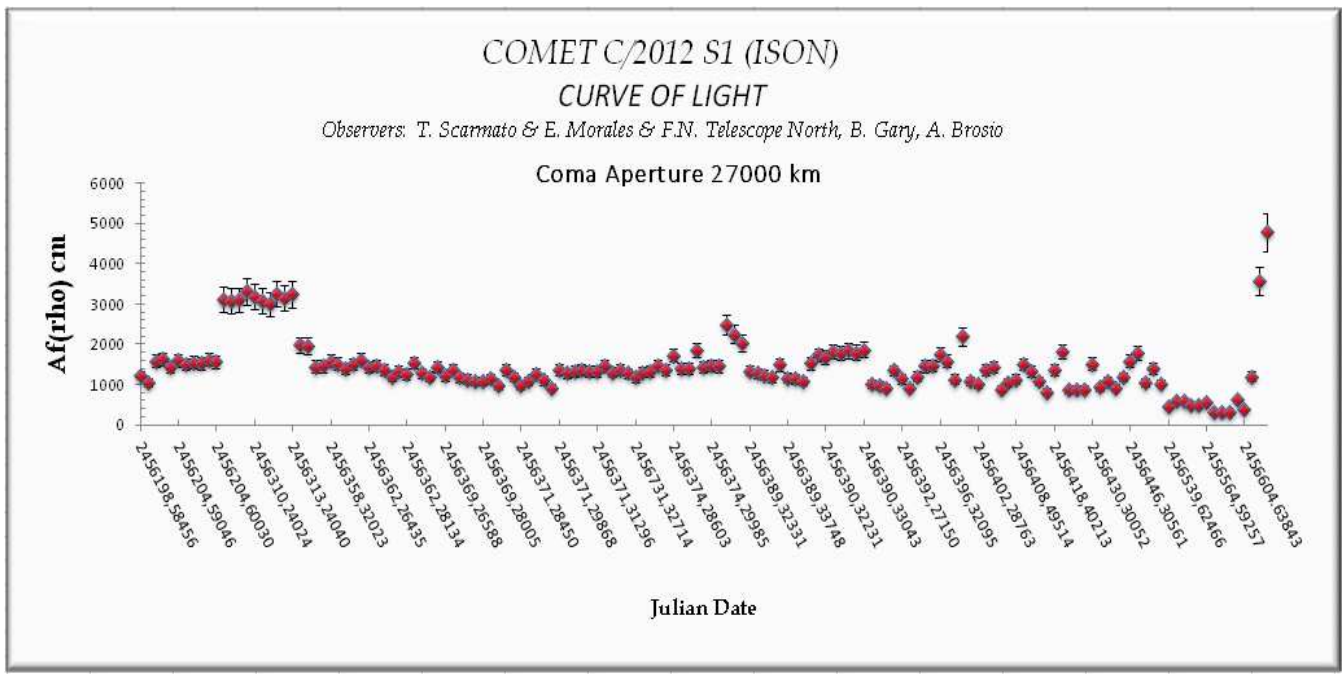

## 4.4 Nucleus radius

| COMET C/2012 S1 (ISON) - 2013 January 21 - Toni Scarmato | | | | | | | | Image from F.T. Telescope North | |
|---|---|---|---|---|---|---|---|---|---|
| Table 1 | ADU | Band R | albedo p=0,04 | | | N. Howes - E. Guido | | | |
| axis | residual | magapp | H (absolute) | Rcomet (m) | err +/- | rho^a | Phase ° | D (AU) | r (AU) |
| x | 551 | 18,428 | 18,348 | 611 | 183 | a=-0,96 | 2 | 4,091 | 5,048 |
| y | 506 | 18,520 | 18,440 | 586 | 175 | a=-0,96 | 2 | 4,091 | 5,048 |
| xy | 942 | 17,846 | 17,766 | 799 | 239 | a=-0,96 | 2 | 4,091 | 5,048 |
| yx | 986 | 17,796 | 17,716 | 818 | 244 | a=-0,96 | 2 | 4,091 | 5,048 |
| Average | 746 | 18,148 | 18,068 | 704 | 200 | a=-0,96 | 2 | 4,091 | 5,048 |
| COMET C/2012 S1 (ISON) - 2013 September 15 - Toni Scarmato's Observatory | | | | | | | | | |
| Table 1 | ADU | Band R | albedo p=0,04 | | | | | | |
| axis | residual | magapp | H (absolute) | Rcomet (m) | err +/- | rho^a | Phase ° | D (AU) | r (AU) |
| x | 123 | 18,353 | 17,583 | 870 | 260 | a=-1,30 | 19,3 | 2,6070 | 1,9420 |
| y | 123 | 18,353 | 17,583 | 870 | 260 | a=-1,30 | 19,3 | 2,6070 | 1,9420 |
| xy | 228 | 17,678 | 16,908 | 1187 | 354 | a=-1,30 | 19,3 | 2,6070 | 1,9420 |
| yx | 223 | 17,707 | 16,937 | 1171 | 350 | a=-1,30 | 19,3 | 2,6070 | 1,9420 |
| Average | 174 | 18,023 | 17,253 | 1024 | 306 | a=-1,30 | 19,3 | 2,6070 | 1,9420 |
| COMET C/2012 S1 (ISON) - 2013 September 15 - Toni Scarmato's Observatory | | | | | | | Image | A. Brosio | |
| Table 1 | ADU | Band R | albedo p=0,04 | | | | | | |
| axis | residual | magapp | H (absolute) | Rcomet (m) | err +/- | rho^a | Phase ° | D (AU) | r (AU) |
| x | 927 | 19,022 | 18,252 | 639 | 191 | a=-1,45 | 19,3 | 2,6070 | 1,9420 |
| y | 889 | 19,067 | 18,297 | 626 | 187 | a=-1,45 | 19,3 | 2,6070 | 1,9420 |
| xy | 1761 | 18,325 | 17,555 | 881 | 263 | a=-1,45 | 19,3 | 2,6070 | 1,9420 |
| yx | 1793 | 18,306 | 17,536 | 889 | 265 | a=-1,45 | 19,3 | 2,6070 | 1,9420 |
| Average | 1343 | 18,680 | 17,910 | 759 | 227 | a=-1,45 | 19,3 | 2,6070 | 1,9420 |

## 4.5 Ison radius compared with other size known comets

**COMET C/2012 S1 (ISON) - 2013 September 4 - Toni Scarmato's Observatory**

| Table 1 axis | ADU residual | Band R magapp | albedo p=0,04 H (absolute) | Rcomet (m) | err +/- | rho^a | Phase ° | D (AU) | r (AU) |
|---|---|---|---|---|---|---|---|---|---|
| x | 84 | 19,217 | 18,625 | 538 | 161 | a=-1,45 | 14,8 | 2,922 | 2,146 |
| y | 86 | 19,192 | 18,600 | 544 | 163 | a=-1,45 | 14,8 | 2,922 | 2,146 |
| xy | 165 | 18,484 | 17,892 | 754 | 225 | a=-1,45 | 14,8 | 2,922 | 2,146 |
| yx | 168 | 18,465 | 17,873 | 761 | 227 | a=-1,45 | 14,8 | 2,922 | 2,146 |
| Average | 126 | 18,840 | 18,248 | 649 | 194 | A=-1,45 | 14,8 | 2,922 | 2,146 |

**COMET C/2011 L4 (PANSTARRS) - 2013 July 29 - Toni Scarmato's Observatory**

| Table 1 axis | ADU residual | Band R magapp | albedo p=0,04 H (absolute) | Rcomet (m) | err +/- | rho^a | Phase ° | D (AU) | r (AU) |
|---|---|---|---|---|---|---|---|---|---|
|  |  |  |  |  | 1AU=149,577,000 km |  |  |  |  |
| x | 71 | 19,682 | 18,830 | 490 | 86 | a=-1,45 | 21,3 | 2,769 | 2,709 |
| y | 70 | 19,698 | 18,846 | 486 | 86 | a=-1,45 | 21,3 | 2,769 | 2,709 |
| xy | 137 | 18,969 | 18,117 | 680 | 120 | a=-1,45 | 21,3 | 2,769 | 2,709 |
| yx | 138 | 18,961 | 18,109 | 682 | 120 | a=-1,45 | 21,3 | 2,769 | 2,709 |
| Average | 104 | 19,328 | 18,476 | 585 | 103 | a=-1,45 | 21,3 | 2,769 | 2,709 |

**COMET 103/P - 2010 August 7 - Toni Scarmato's Observatory**

| Table 1 axis | ADU residual | Band R magapp | albedo p=0,028 H (absolute) | Rcomet (m) | err +/- | rho^a | Phase ° | D (AU) | r (AU) |
|---|---|---|---|---|---|---|---|---|---|
|  |  |  |  |  | 1AU=149,577,000 km |  |  |  |  |
| x | 299 | 19,213 | 18,028 | 848 | 241 | a=-1 | 29,62 | 0,6241 | 1,5086 |
| y | 302 | 19,254 | 18,069 | 832 | 237 | a=-1 | 29,62 | 0,6241 | 1,5086 |
| xy | 520 | 18,512 | 17,660 | 1005 | 286 | a=-1 | 29,62 | 0,6241 | 1,5086 |
| yx | 499 | 18,757 | 17,905 | 898 | 256 | a=-1 | 29,62 | 0,6241 | 1,5086 |
| Average | 405 | 18,934 | 17,916 | 896 | 255 | a=-1 | 29,62 | 0,6241 | 1,5086 |

## 4.6 Sunward Jet Rotation

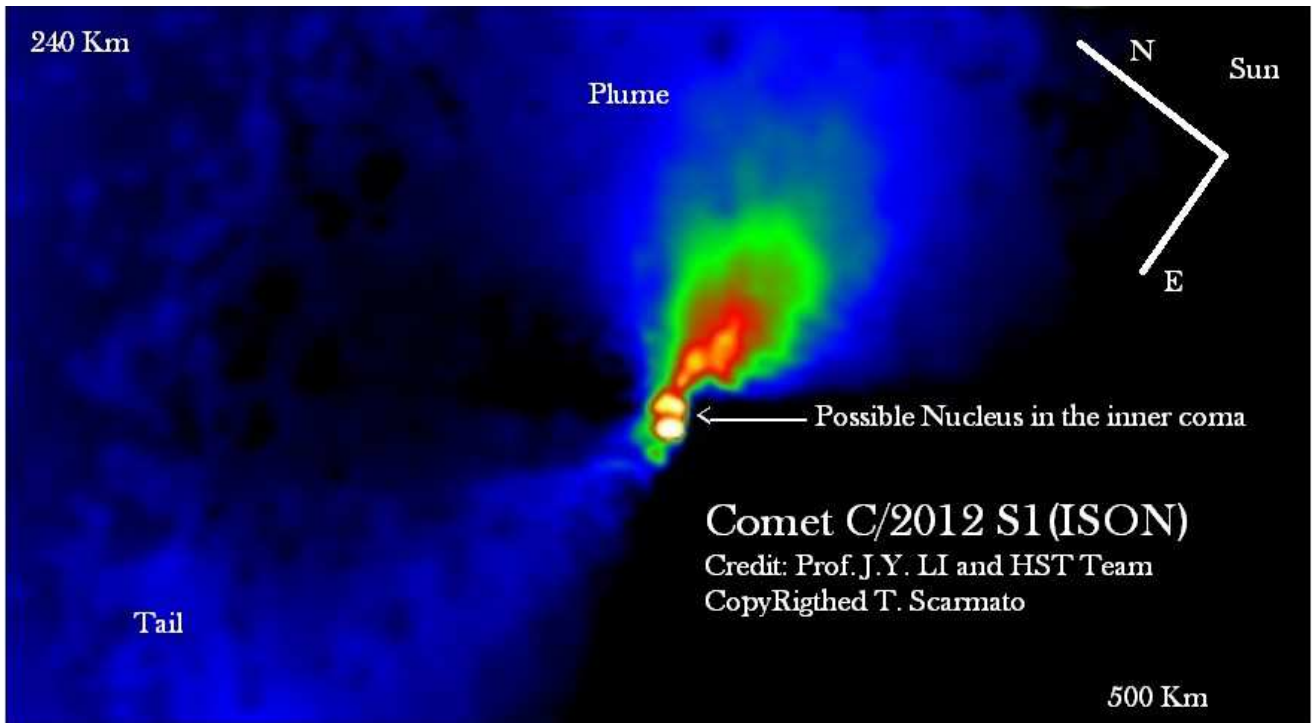

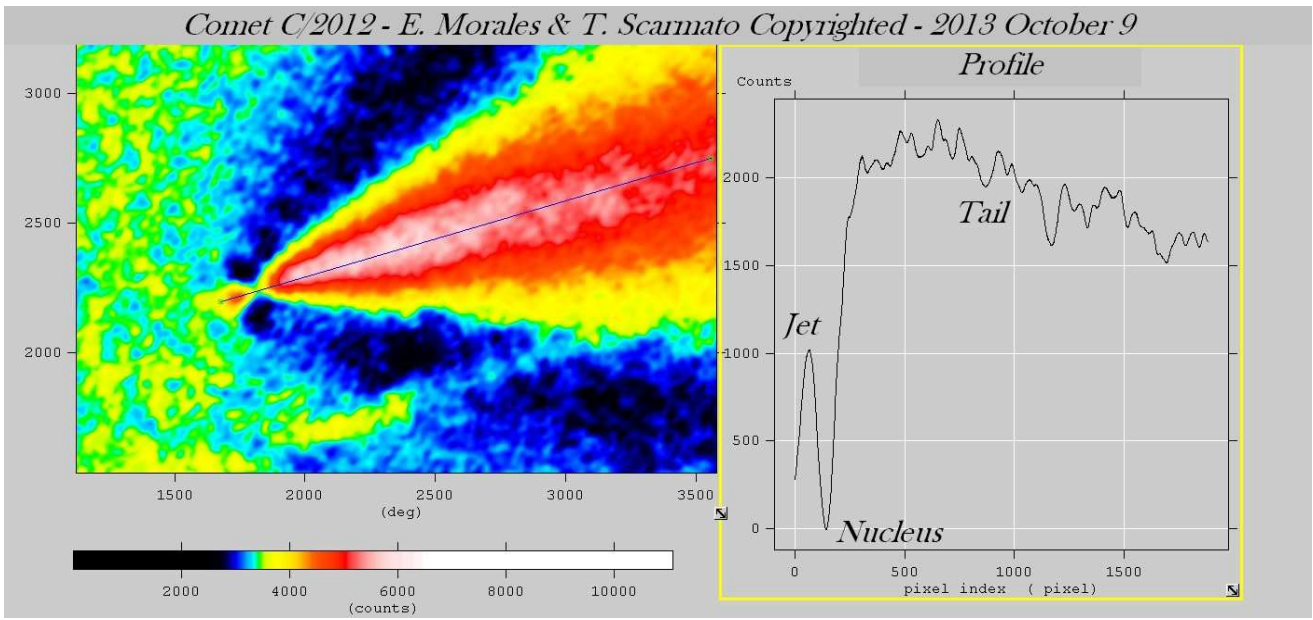

## 2013 October 9 images from Hubble and E. Morales

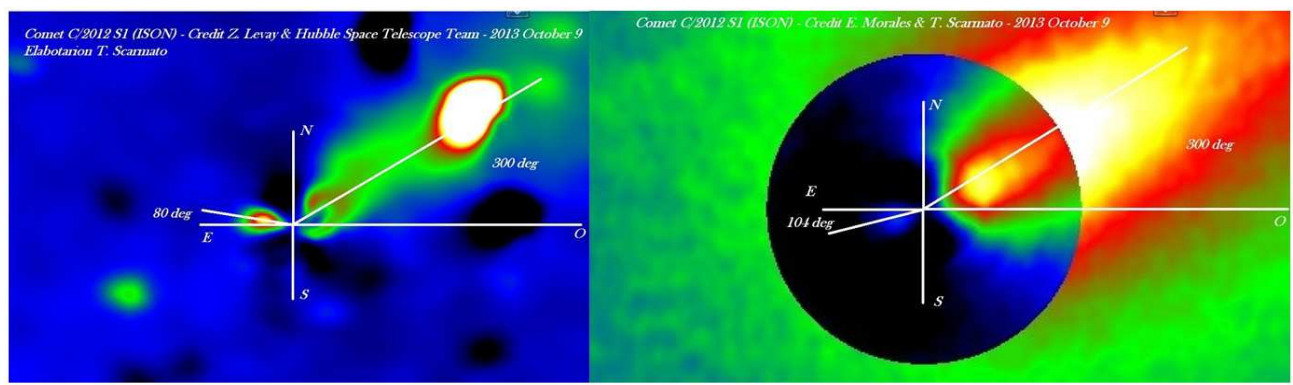

Left image: by Hubble 02:35:30 UT (Larson – Sekanina Filter)

Right image: by E. Morales 09:09:26 UT (MCM Filter)

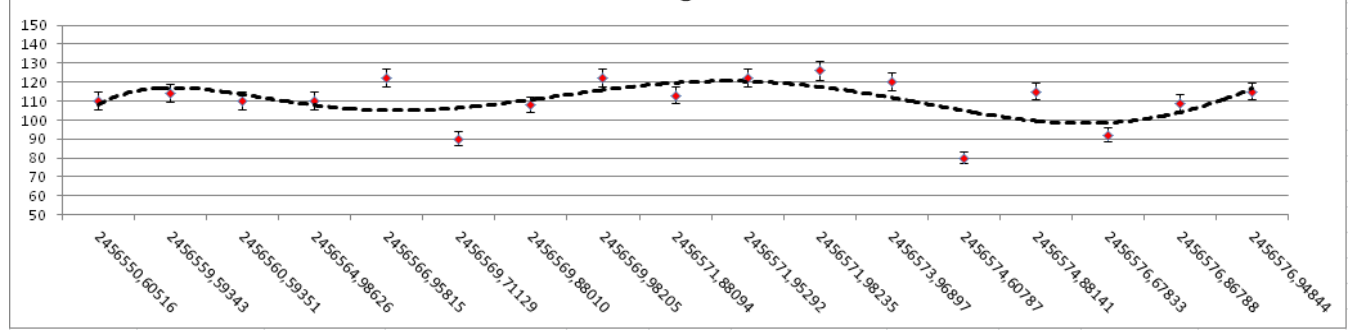

## 4.7 Period computation using Plavchan algorithm

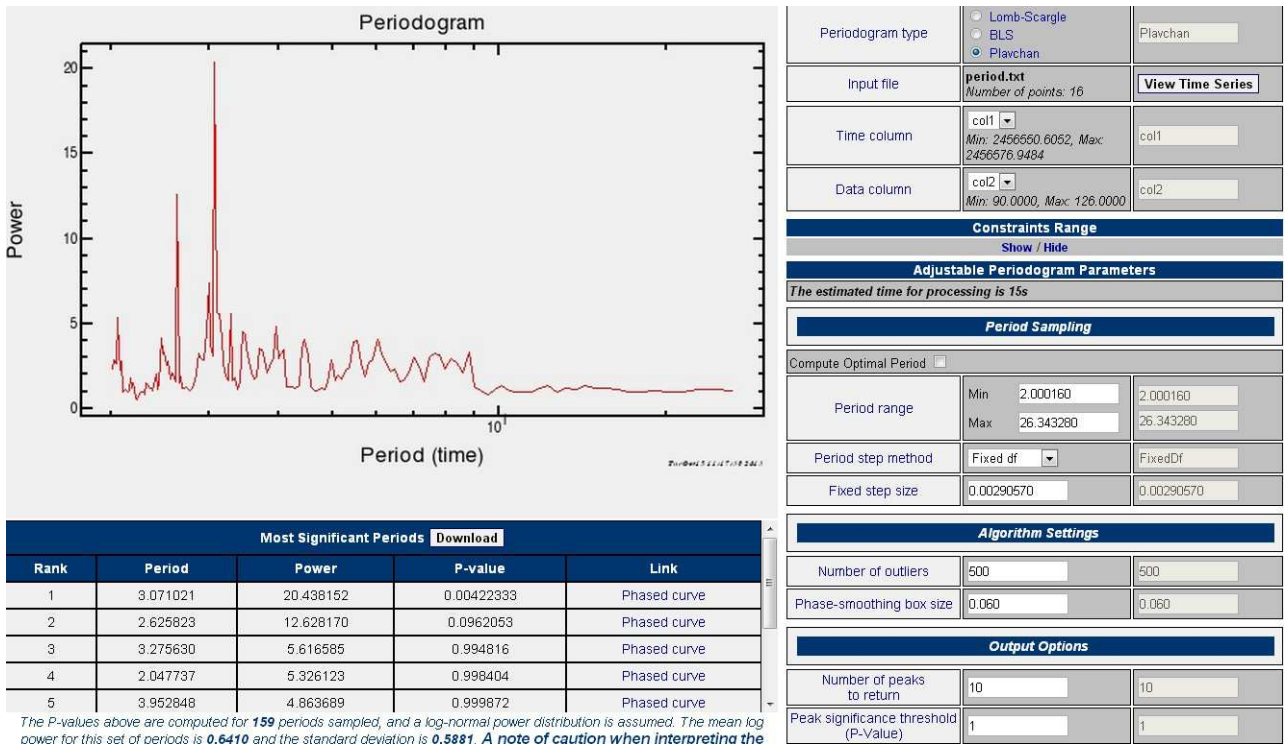

## 4.7 Peculiar structures in the coma and tail and possible fragmentation of the nucleus

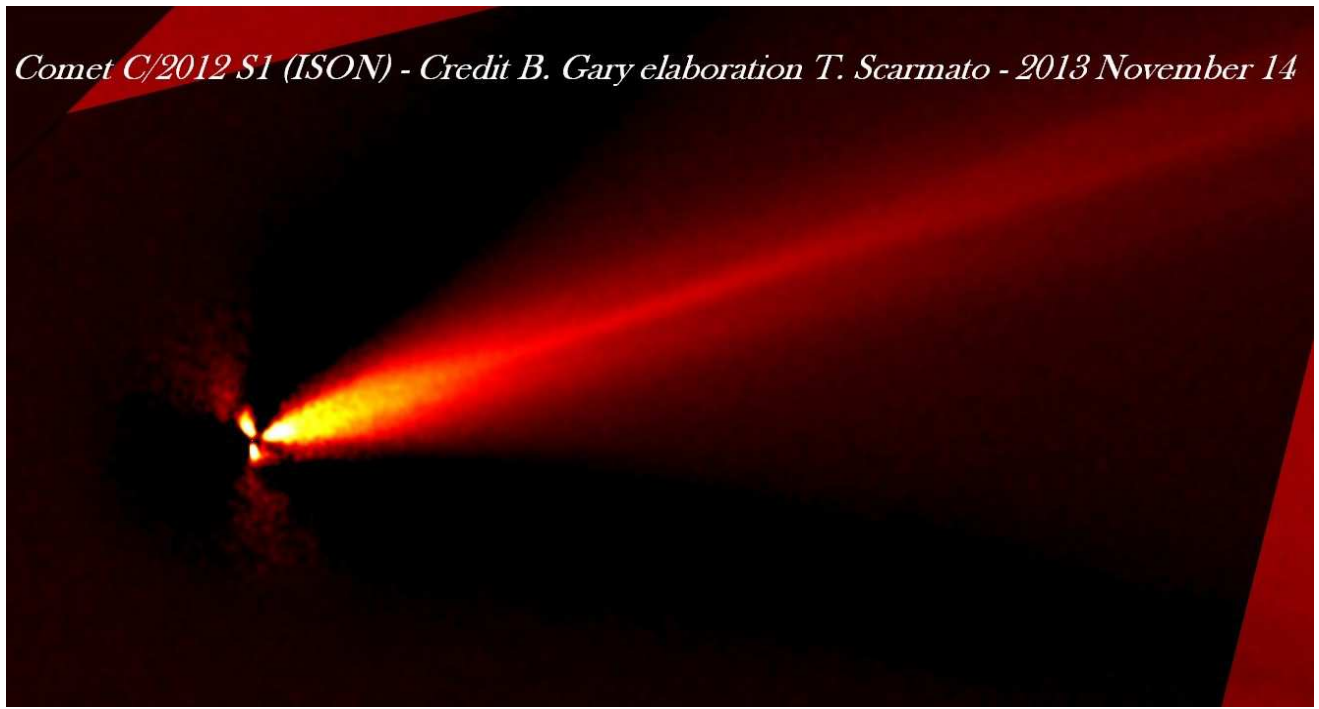

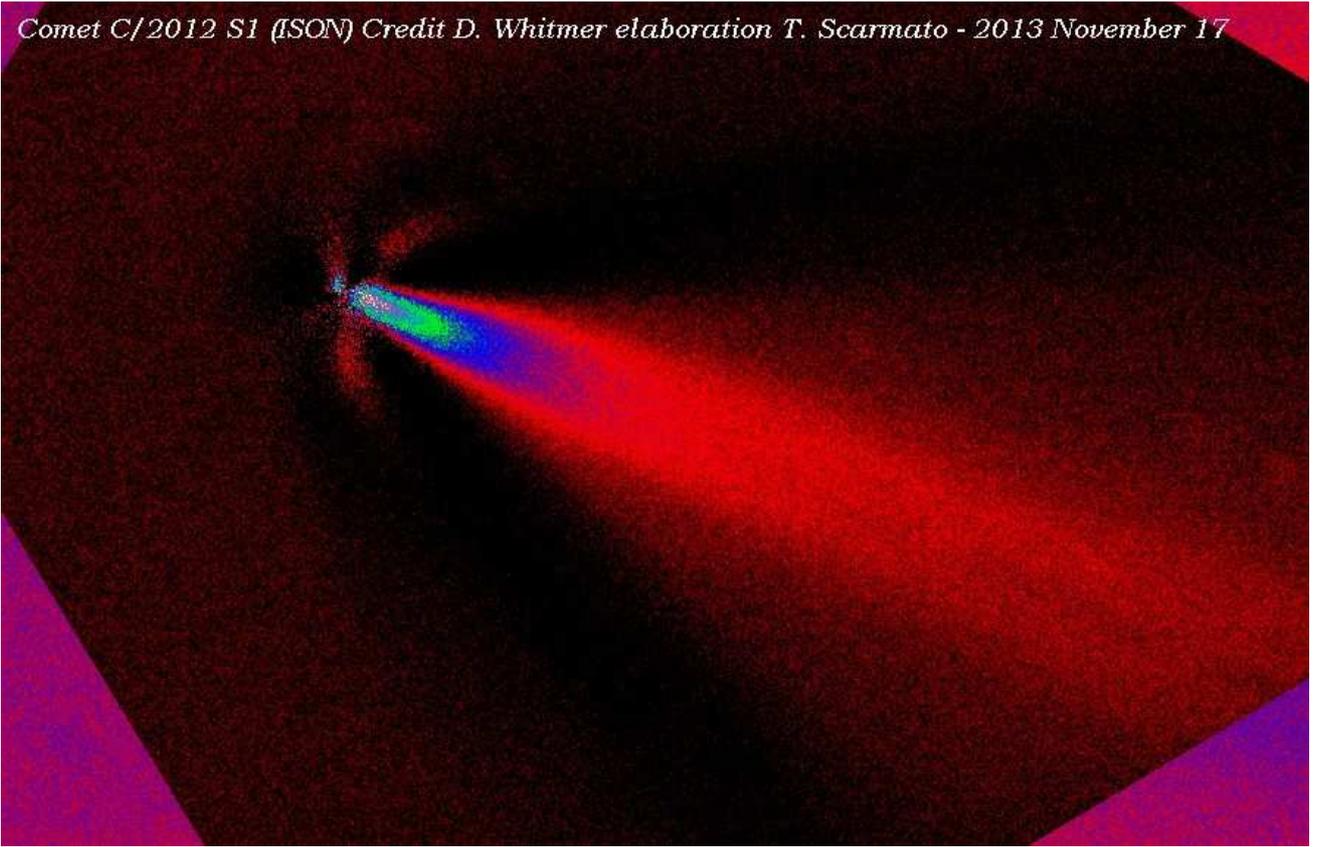

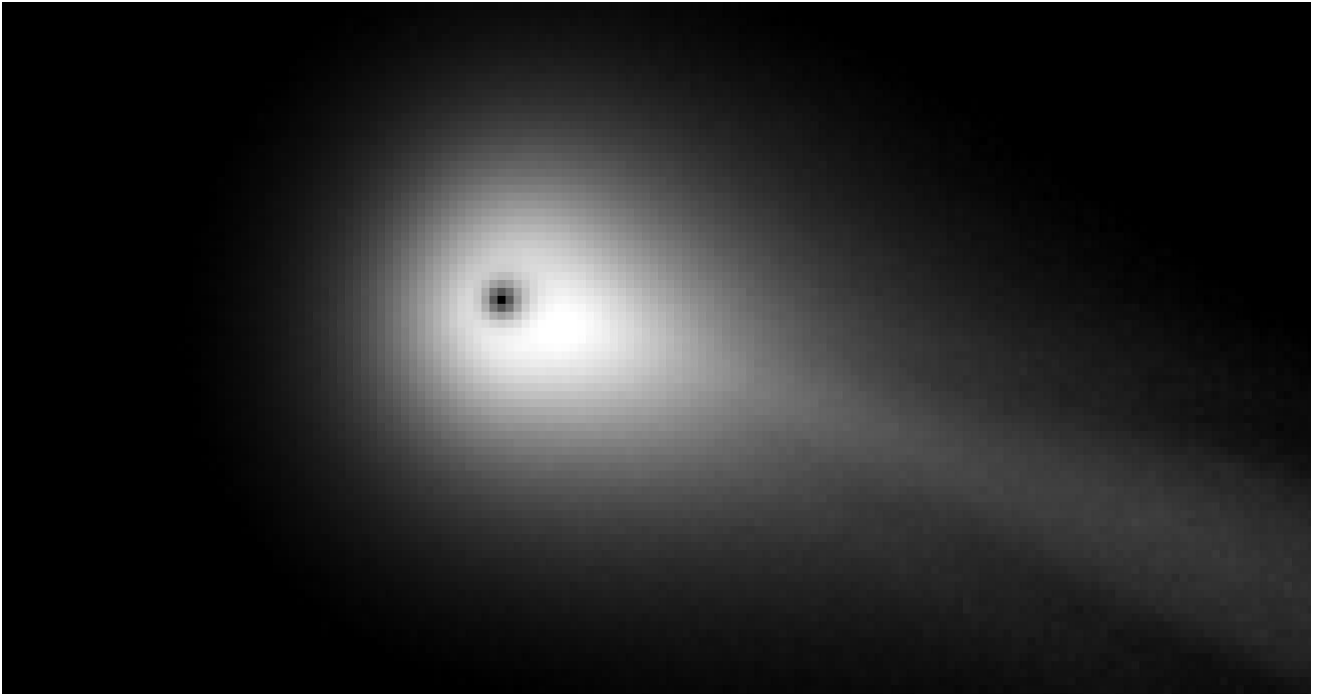

Original image 2013 November 19

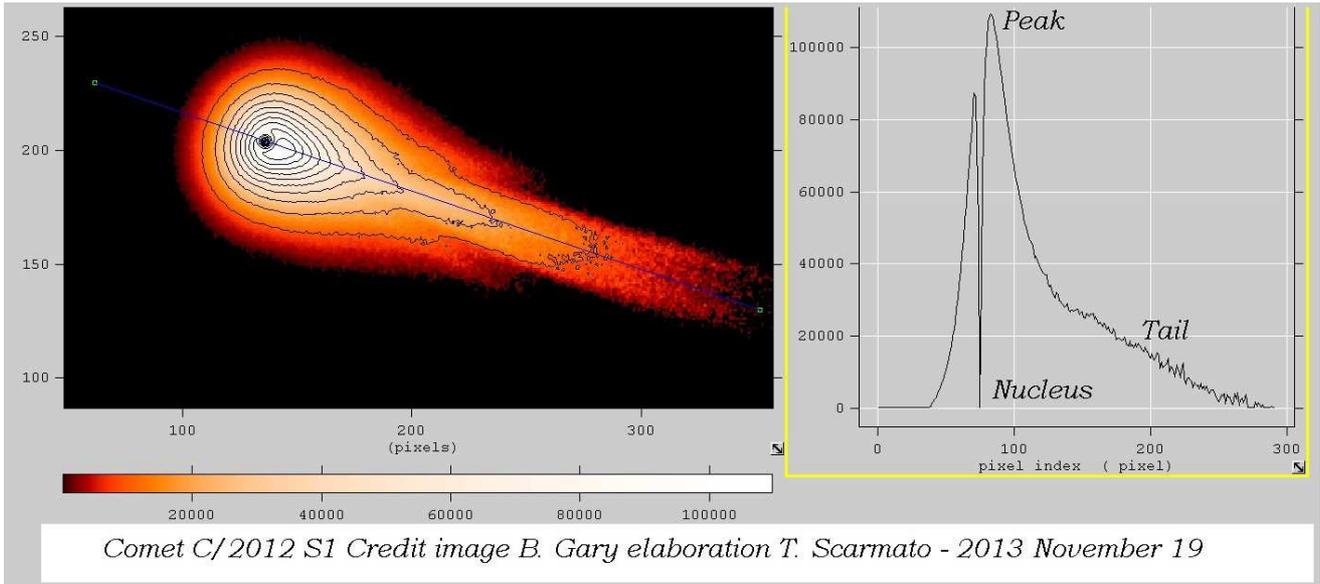

Profile with possible fragment's peak

# Appendix 1

## Basic setup used for observations

$$FOV(') = \frac{3428 * \dim(mm)}{focal(mm)}, \quad scale(arc\sec/pixel) = \frac{FOV}{\dim(pixels)} * 60$$

## Toni Scarmato's Observatory (T. Scarmato Observer)

| Parameters ATIK 16IC mono | |
|---|---|
| Area in Pixel array | 659X494 (325,546 pixels squares) |
| Pixel size | 7.4x7.4 micron |
| Full well depth | 40.000e |
| Dark current | <1e per second at -25°C |
| Peak spectral response | 500 nM |
| Quantum efficiency | >50% a 500 nM |
| A-D converter | 16 bits |
| Readout noise | 7e |
| Anti-blooming | yes |
| Cooling | yes |
| CCD Type | Sony ICX-424AL |
| CCD size (dim area sensitive) | 4,8x3,7 mm |

| Parameters Telescope | | Parameters Filter Rc | |
|---|---|---|---|
| Aperture | 250 mm | Productor | Schuler |
| Focal Lenght | 1200 f/4.8 | Band | Large |
| Scale | 1.27 arcsec/pixel | Lambda peak | 5978 A |
| Optic | Newton | FWHM | 1297 A |
| Type | Reflector | | |
| FOV (Field of View) | 14'x11' | | |

## AOB Observatory (A. Brosio Observer)

| Parameters CCD Camera ATIK 314L+ mono | |
|---|---|
| Area in Pixel array | 1392X1040 (1.447,680 pixels squares) |
| Pixel size | 6.45x6.45 micron |
| Full well depth | 175.000e |
| Dark current | <1e per second at -15°C |
| Peak spectral response | 500 nM |
| Quantum efficiency | >65% a 540 nM |
| A-D converter | 16 bits |
| Readout noise | 7e |
| Anti-blooming | yes |
| Cooling | yes |
| CCD Type | Sony ICX-285 |
| CCD size (dim area sensitive) | 10.2mm x 8.3 mm |

| Parameters Telescope | | | Parameters Filter Rc | |
|---|---|---|---|---|
| Aperture | 250 mm | | Productor | Schuler |
| Focal Lenght | 1200 f/4.8 | | Band | Large |
| Scale | 1.10 arcsec/pixel | | Lambda peak | 5978 A |
| Optic | Newton | | FWHM | 1297 A |
| Type | Reflector | | | |
| FOV (Field of View) | 25'x19' | | | |

## Acknowledgements.

E. Morales; N. Howes; A. Brosio; B. Gary; D. Whitmer; G. Muler; J.A. Jimenez; Seeichi Yoshida for useful program Comet for Win; R. Bidoli for useful help; **Senior Research Scientist** Padma Yanamandra-Fisher (Space Science Institute); **Senior Research Scientist** Jian Yang-Li (Planetary Science Institute); **Image Group Lead** Z.G. Levay Nasa/Esa Hubble Team, (Space Telescope Science Institute); for the useful hints and the permission to use Hubble's images.

______________________________